# Wind-invariant saltation heights imply linear scaling of aeolian saltation flux with shear stress

Raleigh L. Martin*, Department of Atmospheric and Oceanic Sciences, University of California, Los Angeles, CA 90095

Jasper F. Kok, Department of Atmospheric and Oceanic Sciences, University of California, Los Angeles, CA 90095

*raleighm@atmos.ucla.edu

**Abstract**
Wind-driven sand transport generates atmospheric dust, forms dunes, and sculpts landscapes. However, it remains unclear how the sand flux scales with wind speed, largely because models do not agree on how particle speed changes with wind shear velocity. Here, we present comprehensive measurements from three new field sites and three published studies, showing that characteristic saltation layer heights, and thus particle speeds, remain approximately constant with shear velocity. This result implies a linear dependence of saltation flux on wind shear stress, which contrasts with the nonlinear 3/2 scaling used in most aeolian process predictions. We confirm the linear flux law with direct measurements of the stress-flux relationship occurring at each site. Models for dust generation, dune migration, and other processes driven by wind-blown sand on Earth, Mars, and several other planetary surfaces should be modified to account for linear stress-flux scaling.

# Introduction

Understanding wind-driven ('aeolian') sand transport has been critical to modeling a wide range of geophysical processes. In models of coastal evolution and the effects of sea-level rise, aeolian sand flux contributes to formation of protective foredunes *(1)*. In arctic and alpine environments, aeolian snow saltation *(2)* modulates snow depth and melting processes *(3)*. In desert and semiarid environments, aeolian sand transport drives evolution of complex dune fields *(4, 5)*, abrasion of bedrock *(6)*, erosion of soil *(7)*, and generation of atmospheric mineral dust *(8, 9)*. These dust aerosols have numerous important effects on the Earth system, including nutrient fertilization for land and ocean biota *(10)*, modification of the hydrological cycle *(11)*, and alteration of the Earth's climate by scattering and absorbing radiation and seeding clouds *(12)*. Further, aeolian transport models, used in conjunction with observations of ripple and dune migration, are now allowing the inference of atmospheric conditions on Mars and other planetary bodies *(13–15)*.

Despite the importance of aeolian transport to this wide range of processes on Earth and other planetary bodies, existing aeolian transport models produce results that are inconsistent with each other *(16)*. For a given wind stress, these models predict substantially different sand fluxes based on differing understandings of saltation, the ballistic hopping of sand-sized particles driven by the wind. Even though models agree on the core particle-wind interactions (e.g., particle collisions, drag) producing saltation, they treat these processes differently and thereby produce widely varying predictions of sand flux. The most fundamental resulting difference in saltation model predictions is the scaling relationship between sand flux $Q$ and wind stress $\tau$, i.e. $Q \sim \tau^f$, where $f$ is the flux scaling exponent. Beginning from the classic work of Bagnold, most saltation models predict $f = 3/2$ nonlinear flux scaling *(17, 18)*. Such 3/2 models are prevalent for applications including prediction of atmospheric dust emission *(19, 20)*, dune migration *(21, 22)*, and planetary surface evolution *(13)*. More recent models scale sand flux linearly ($f = 1$) *(23–25)*, or weakly nonlinearly between the linear and 3/2 end-member cases ($1 < f < 3/2$) *(26, 27)*. Alternatively, in terms of shear velocity $u_*$ ($= \sqrt{\tau/\rho_f}$, where $\rho_f$ is air density), saltation scaling predictions range from squared ($Q \sim u_*^2$) to cubic ($Q \sim u_*^3$).

The controversy in saltation flux scaling originates from disagreement over particle speed scaling. The total saltation flux $Q$ [gm$^{-1}$] equals the product of the vertically integrated saltation layer mass concentration $\Phi$ [gm$^{-2}$] and the mean horizontal particle speed $V$ [ms$^{-1}$]:

$$Q = \Phi V. \tag{1}$$

Thus, the saltation flux scaling with wind shear stress ($Q \sim \tau^f$) combines concentration scaling ($\Phi \sim \tau^c$) and particle speed scaling ($V \sim u_*^r$), so that roughly $f = c + r/2$ by equation 1. Models and observations agree that mass concentration scales linearly with shear stress (i.e. $c \approx 1$) in excess of a minimum threshold *(26, 28)*, typically the "impact threshold" stress $\tau_{it}$ required to sustain saltation *(16, 18, 29, 30)*. However, models disagree on whether particle speed scales linearly with $u_*$, i.e. $r = 1$ *(17, 18, 31)*, or remains roughly constant with $u_*$, i.e. $r \approx 0$ *(23, 28)*. The classic $r = 1$ models, which imply $f = 3/2$ nonlinear flux scaling, assume that saltation trajectories are initiated primarily by fluid lifting, whereas more recent $r \approx 0$ models treat particle entrainment as dominated by ejection ('splash') *(16)*. In support of these recent $r \approx 0$ models, wind tunnel observations *(28, 32–35)* show that near-surface particle speeds $v_0$ do not change with $u_*$, which is possible only for splash-dominated entrainment *(16)*. Based on this evidence for constant $v_0$, recent models *(16, 23–25, 28, 36)* also assume constant $V$ ($r = 0$) and therefore linear flux scaling ($f = 1$), though other models *(26, 27)* allow a weak increase in $V$ with $u_*$ ($0 < r < 1$) and weakly nonlinear saltation flux scaling ($1 < f < 3/2$). The few existing field studies that have sought to evaluate these saltation models *(37, 38)* have lacked the statistical precision to test different flux laws, and none have directly addressed particle speed.

In this paper, we offer resolution of this continuing controversy over the scaling of aeolian saltation flux with wind speed. We do so using comprehensive field-based measurements of saltation at multiple distinctive sites. We show that saltation layer height, and thus particle speed, does not change with increasing wind shear velocity. We then confirm the linear scaling between wind stress and saltation flux implied by these constant saltation heights using direct field measurements of sand flux. Finally, we discuss the implications of these results for representing aeolian processes on Earth, Mars, and other planetary bodies.

**Table 1. Grain size, saltation profile, and flux law fit values.**

| Location / study | $d_{50}$ (mm) | $\langle z_q \rangle$ (m) | $\tau_{it}$ (Pa) | $u_{*,it}$ (m/s) | $C_Q$ |
|---|---|---|---|---|---|
| Jericoacoara | 0.53 ± 0.04 | 0.097 ± 0.005 | 0.135 ± 0.015 | 0.341 ± 0.019 | 7.3 ± 0.9 |
| Rancho Guadalupe | 0.53 ± 0.03 | 0.107 ± 0.005 | 0.110 ± 0.021 | 0.300 ± 0.028 | 5.8 ± 0.5 |
| Oceano | 0.40 ± 0.07 | 0.055 ± 0.004 | 0.094 ± 0.006 | 0.277 ± 0.009 | 5.9 ± 1.0 |
| Greeley et al. (1996)*(40)* | 0.23 | 0.050 ± 0.006 | | | |
| Namikas (2003)*(41)* | 0.25 | 0.049 ± 0.003 | | | |
| Farrell et al. (2012)*(42)* | | 0.081 ± 0.008 | | | |

$d_{50}$ is median grain diameter for surface samples. Full grain size distributions for field sites can be found in Martin et al. *(43)*. Farrell et al. *(42)* did not report a surface grain size, and Greeley et al. *(40)* and Namikas *(41)* did not report associated uncertainties. $\langle z_q \rangle$ is mean saltation layer height. $\tau_{it}$ and $u_{*,it}$ are best fit saltation impact threshold shear stress and shear velocity, respectively, for a linear flux law. $C_Q$ is the best fit scaling parameter for equation 3. Included uncertainties here and elsewhere represent one standard deviation.

## Results
*Determining saltation layer heights as a proxy for particle speed scaling*
We measured time series of wind velocity and vertical profiles of streamwise saltation flux at three coastal sand dune locations with varying site conditions: Jericoacoara (Brazil), Rancho

Guadalupe (California), and Oceano (California). Wind and sand transport variables were computed over 30-minute time intervals (see Materials and Methods), sufficiently long to capture the full range of driving turbulent fluctuations *(39)*. We also analyzed field saltation profile measurements obtained by Greeley et al. *(40)*, Namikas *(41)*, and Farrell et al. *(42)* (see supplementary text). Taken together, our new field data and the literature data represent a wide range of surface sand sizes (Table 1).

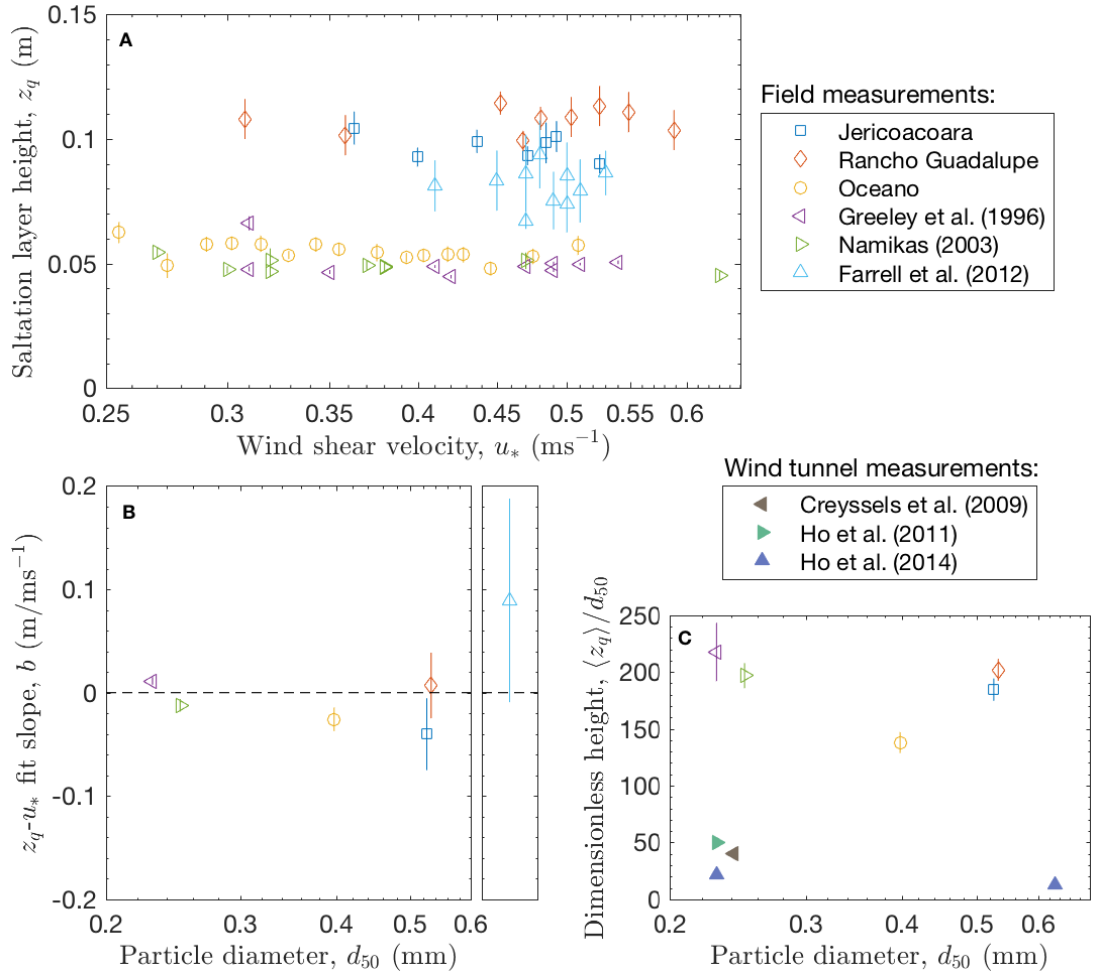

**Fig. 1. Measurements of saltation layer heights.** (**A**) Characteristic saltation layer heights $z_q$ versus shear velocities $u_*$, acquired over 30 minute intervals, grouped into $u_*$ bins. Bars denote uncertainties in $z_q$ for each bin. Methods for computing $z_q$ from field data of Greeley et al. *(40)*, Namikas *(41)*, and Farrell et al. *(42)* are described in the supplementary text. (**B**) Slope parameter $b$ for linear fit to $z_q = a + bu_*$, versus $d_{50}$. Data are plotted separately for Farrell et al. *(42)*, who did not report $d_{50}$. Bars denote uncertainties in $b$ for each site. (**C**) Mean dimensionless saltation layer height $\langle z_q \rangle / d_{50}$ versus particle diameter $d_{50}$. Bars denote uncertainties in $\langle z_q \rangle / d_{50}$ for each site. Dimensionless saltation layer heights from wind tunnel experiments *(28, 35, 36)* are shown for comparison.

As a proxy for particle speed scaling, we examined the relationship between the characteristic saltation layer decay height $z_q$ and shear velocity $u_*$. As direct measurements of particle speed remain difficult to obtain under natural field conditions *(40)*, we instead relied on the fact that $z_q$ scales with particle speed squared *(18)* (see Materials and Methods). To determine $z_q$, we fit an exponential function to vertical profiles of partial (height-specific) saltation flux, $q(z)$ [gm$^{-2}$]:

$$q(z) = q_0 \exp\left(-\frac{z}{z_q}\right), \qquad (2)$$

where $q_0$ [gm$^{-2}$] is the scaling parameter for the profile. Our choice to fit an exponential to the flux profile was justified by previous wind tunnel observations *(36)* and by its close adherence to measured flux profiles *(43)*.

We find that $z_q$ remains roughly constant with $u_*$ at each field site (Fig. 1A). The same is true for calculations of saltation layer height obtained by fitting equation 2 to the field measurements of Greeley et al. *(40)*, Namikas *(41)*, and Farrell et al. *(42)* (supplementary text). Though substantial differences in saltation layer height exist from site to site, the slopes of the linear fits to $z_q$ versus $u_*$ (Fig. 1B) show that changes in saltation layer height with shear velocity are statistically insignificant or negligible. When $z_q$ is normalized by median grain diameter $d_{50}$, variability among sites is reduced substantially, with mean dimensionless saltation layer heights, $\langle z_q \rangle / d_{50}$, all falling within the range of 138–218 (Fig. 1C and Table 1). Our measurements therefore show that $z_q \propto d_{50}$, regardless of $u_*$. Since changes in saltation layer height with shear velocity appear negligible, our data indicate that other site conditions, primarily $d_{50}$, exert the dominant control on particle trajectories.

As derived in previous work *(16)*, the scaling of saltation height with particle diameter yields a simple scaling of saltation flux with wind shear stress:

$$Q = C_Q \frac{u_{*,it}}{g} \tau_{ex}, \qquad (3)$$

where $C_Q$ is an empirically-derived dimensionless flux scaling parameter, $u_{*,it}$ is the impact threshold shear velocity, $g$ is gravitational acceleration ($\approx$ 9.8 ms$^{-2}$ at Earth sea level), and $\tau_{ex} = \tau - \tau_{it}$ is the "excess" stress available to do work to move sediment. A full derivation of equation 3 is provided in Materials and Methods.

*Evaluation of linear saltation flux law using direct field measurements*
To evaluate the flux law prediction of equation 3, we first estimate $\tau_{it}$ for each field site from the zero-intercept of linear fit to $Q$ versus $\tau$ (Table 1; see also Materials and Methods and Fig. S2). Then we compute total flux as $Q = q_0 z_q$ based on the exponential profile fit values (equation 2). We find that, at all field sites, $Q$ is roughly proportional to $\tau_{ex}$ (Fig. 2), thus supporting a linear flux law.

To rule out the possibility of a nonlinear flux law, for each field site we compare the linear fit to a nonlinear 3/2 best fit of the form $Q \sim u_* \tau_{ex}$ (see Materials and Methods and supplementary text). At Jericoacoara and Oceano, the mean-squared difference between best fit and observed values is substantially smaller for the linear fit than for the nonlinear 3/2 fit. At Rancho Guadalupe, the mean-squared differences are comparable (Fig. S2 and Table S1).

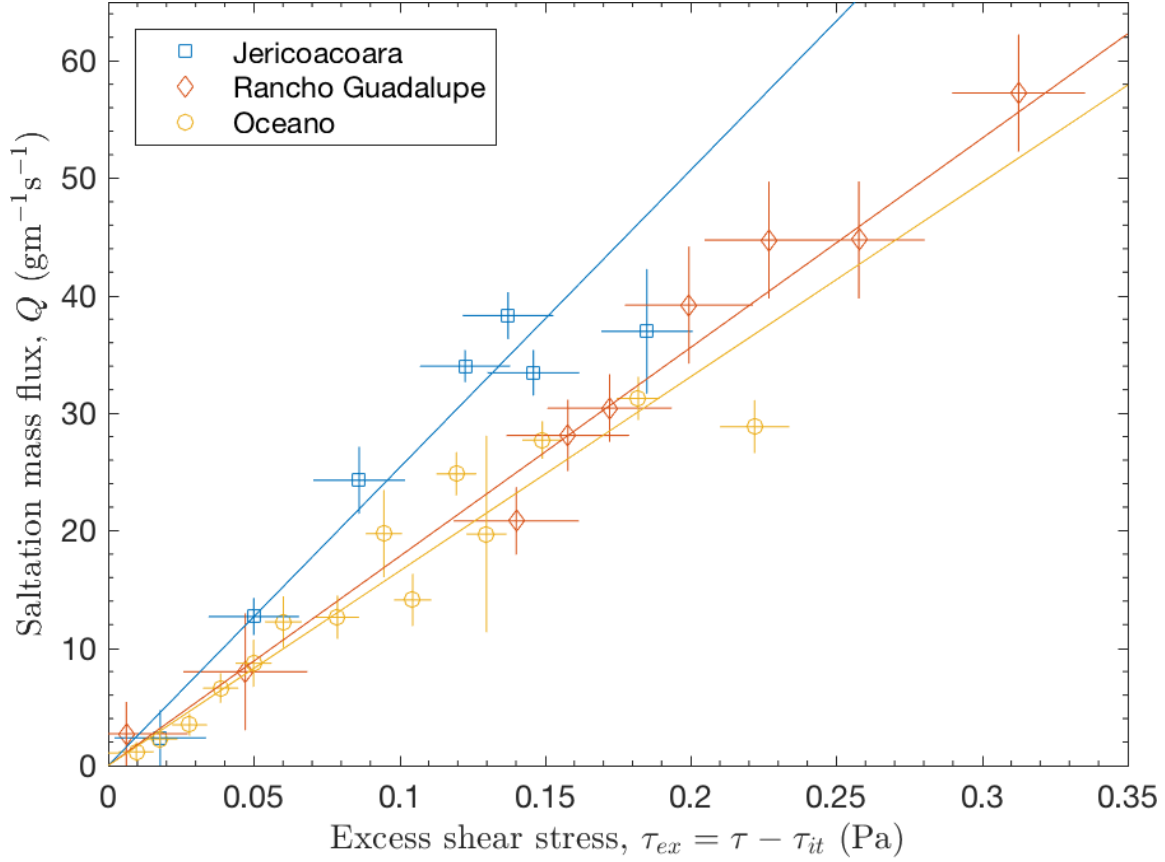

**Fig. 2. Saltation mass flux $Q$ versus excess shear stress $\tau_{ex}$.** Fluxes are grouped into $\tau_{ex}$ bins with vertical bars denoting flux uncertainties and horizontal bars denoting stress uncertainties for each bin. The impact threshold stress $\tau_{it}$ for calculating $\tau_{ex}$ was computed separately for each site (Table 1 and Fig. S2). Solid lines indicate predictions for $Q$ obtained by equation 3 with flux scaling parameter $C_Q$ taken as mean estimated value for this parameter ($\hat{C}_Q$, equation 4) calculated for each site.

---

To examine the variations in saltation flux scaling from site to site, we obtain estimates of the dimensionless flux scaling parameter $\hat{C}_Q$ by rearranging equation 3:

$$\hat{C}_Q = \frac{g}{u_{*,it}} \frac{Q}{\tau_{ex}}. \tag{4}$$

Fig. 3 shows that $\hat{C}_Q$ remains roughly constant with excess shear stress for each site. The constancy of $\hat{C}_Q$ further supports the linear form of the flux law, and it indicates that $u_{*,it}$ accounts for most of the variability in sand flux among sites. The mean values of $\hat{C}_Q$ for each site fall in the approximate range of 5.8 to 7.3 (Table 1). Based on the finding of a constant $\hat{C}_Q$, and restating equation 3 in its conventional form with shear velocity, we obtain:

$$Q = C_Q \frac{u_{*,it}}{g} \left( u_*^2 - u_{*,it}^2 \right), \tag{5}$$

with $C_Q = 6.1 \pm 0.4$. Notably, our empirically obtained flux scaling parameter closely matches the $C_Q = 5$ predicted by Kok et al. *(16)* in their derivation of equation 3 based on typical observed saltator hop lengths, particle speeds, and impact threshold shear velocities for fine sand. This agreement lends support to the physical interpretations and assumptions underlying derivation of the linear saltation flux law (see Materials and Methods).

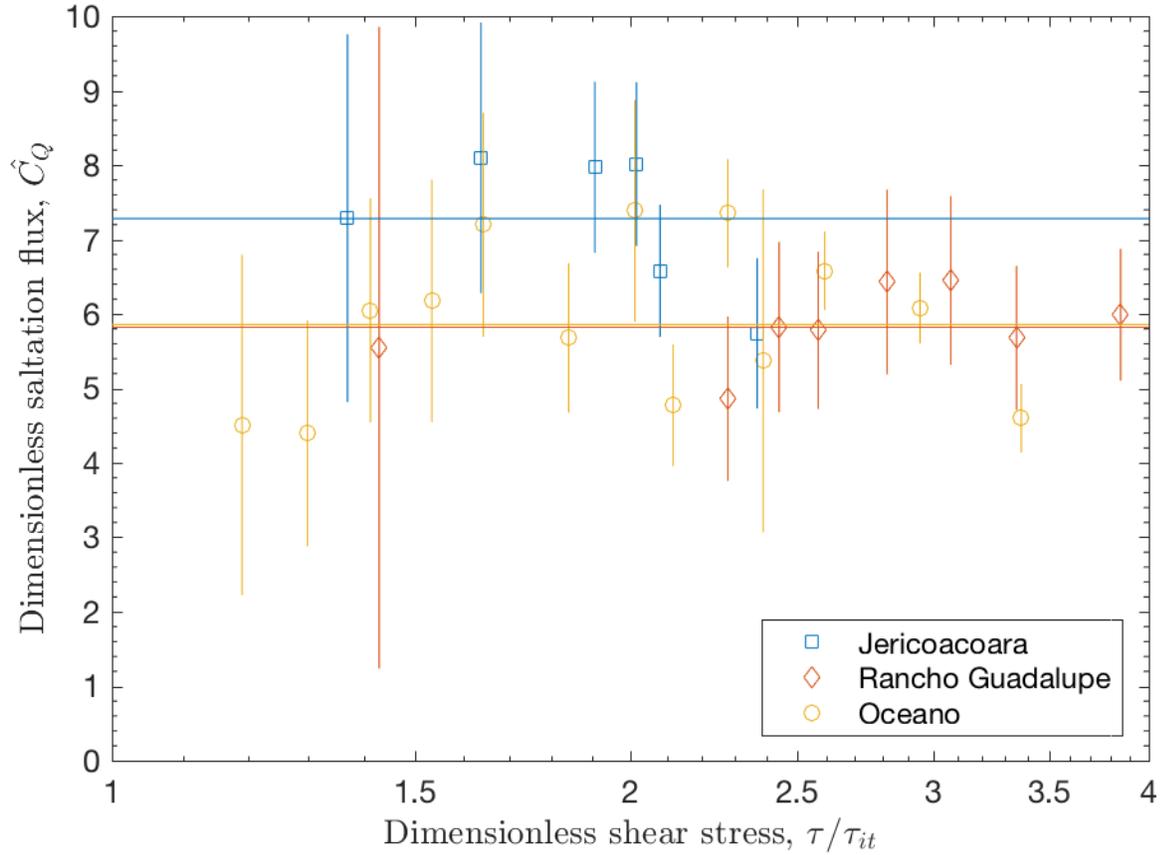

**Fig. 3. Dimensionless saltation flux versus dimensionless shear stress.** Estimated dimensionless saltation flux scaling parameter $\hat{C}_Q$ is computed using equation 4, and dimensionless shear stress is calculated as $\tau/\tau_{it}$. Data are grouped into $\tau/\tau_{it}$ bins with bars denoting $\hat{C}_Q$ uncertainties for each bin. The mean values of $\hat{C}_Q$ for each site, denoted by horizontal lines, all fall in a similar range (Table 1), supporting the linear flux law described in equation 3.

## Discussion

Our results shed light on one of the most important controversies in aeolian science: how does the sand flux scale with wind speed? To address the core of this controversy – the relationship between particle and wind speeds – we presented field data from six distinctive sites indicating that saltation layer height does not change with shear velocity (Fig. 1). This finding implies that mean particle speeds remain constant with shear velocity. Therefore, saltation flux changes solely through changes in particle concentration, which increases linearly with excess shear stress (equation 3). We confirmed the resulting linear scaling of saltation flux with wind stress using field measurements at three sites with distinct soil size distributions (Fig. 2). From these measurements, we then obtained a specific parameterization for the flux relationship (Fig. 3) and a simple equation (equation 5) for modeling saltation flux. This work represents the first strong field-based evidence for the linear flux law, thereby overturning widely-used nonlinear flux models that incorrectly assume particle speeds to increase with shear velocity.

Even though previous field studies have been performed to evaluate flux laws, they have generated insufficient measurements to overcome the wide variability of natural saltation. Our field campaigns have addressed this problem with measurements covering longer observational windows (30 min here versus 17 min in Sherman and Li *(38)*) and including a greater number of time intervals (154 here versus 51 in Sherman and Li *(38)*). Longer observational windows capture the full range of large-scale turbulence structures *(44, 45)*. Further, the extensiveness of our data sets allowed us to group our data into bins corresponding to specific ranges of shear stress (see Materials and Methods and supplementary text). In our measurements, fluxes typically varied by $\approx$ 50% within each $\tau$ bin, and these variations were even larger near the saltation threshold $\tau_{it}$. Our approach enabled us to average the measurements from multiple time intervals within each bin, thereby reducing the substantial noise apparent among comparisons of individual observations (Figs. 7 and 8 in Martin et al. *(43)*).

We consider four potential limitations of our analysis. First, the proportionality constant $C_Q$ in our linear flux law (equation 3) varies by $\approx$10–20% both within and among sites (Fig. 3 and Table 1). Although this variability is substantially smaller than the scatter of measurements around nonlinear flux laws shown by previous studies *(38)*, variations in $C_Q$ imply that there remain unexplained sources of variability in saltation flux. Whereas equation 3 is derived by assuming a constant bulk restitution coefficient *e*, it is possible that soil moisture *(46)* and grain-size distributions *(47)* cause variations in *e*. Though these variations account for only a small fraction of variability among field measurements, future saltation models could address them by directly parameterizing *e* in the derivation of the flux law (see Materials and Methods). Second, our observations supporting a linear flux law cover a limited range of shear stresses ($\tau/\tau_{it} = 0$–4). Durán et al. *(48)* suggested that flux scaling transitions from linear to nonlinear when $\tau/\tau_{it} > 2$. However, we find a slight *reduction* in the scaling rate when $\tau/\tau_{it} > 2$, which contrasts with the Durán et al. prediction. Third, our measurements cover the upper range of the flux profile ($z > 2$ cm). Namikas *(41)* and Bauer and Davidson-Arnott *(49)* observed saltation fluxes very close to the surface ($z < 2$ cm) that exceed the expectation of an exponential profile (equation 2). The fact that we did not measure this lower range likely resulted in an underestimation of the total flux. However, profile fits to the Namikas data, which do extend to below $z = 2$ cm, show no systematic change in $z_q$ with $u_*$, suggesting that this lower region contributes a constant fraction of the total flux. Thus, including this flux close to the surface would be unlikely to change our observation of linear flux scaling. Fourth, our measurements of saltation flux may contain flux-dependent biases resulting from variations in saltation trap efficiency *(37, 50)*. Such biases could produce height-dependent relative errors in the profile for $q(z)$ and thus systematic errors in $z_q$ and $Q$. However, different types of traps (with different expected biases) were used by Greeley et al., Namikas, Farrell et al., and at our field sites, yet all produced roughly constant $z_q$ with $u_*$. It is thus unlikely that trap bias affected our result of constant saltation layer heights.

Despite these potential limitations, the remarkable constancy of saltation layer height with shear velocity (and implied constancy of particle speed $V$) across six field sites offers answers to lingering questions about saltation process mechanics. Our result that $V$ remains constant with $u_*$ in natural saltation supports past wind tunnel observations of constant near-surface particle speeds $v_0$ *(28, 32–35)* and the implication that particle entrainment is dominated by splash *(23, 25, 36, 48, 51)*. Furthermore, these results suggest that particle speed profiles away from the bed $v(z > 0)$ remain unchanged with $u_*$. However, such constancy of $v(z > 0)$ conflicts with recent simulations *(48, 52)* and wind-tunnel experiments *(28, 33, 36)* finding that particle speeds away from the bed do in fact increase with $u_*$ (Fig. 14 in Kok et al. *(16)*). To accommodate such increasing particle speeds, $z_q$ must also increase with $u_*$ *(16)*; however, this conflicts with wind tunnel measurements *(28, 35, 36)* and our field observations showing constant $z_q$.

We also observed normalized saltation heights of $z_q/d_{50} \approx 150$–$200$ (Fig. 1C) that are several times larger than the $z_q/d_{50} \approx 10$–$50$ obtained in equilibrated wind tunnel *(28, 35, 36)* and some numerical *(53)* studies, indicating a gap in our current understanding of saltation mechanics informed by experiments and simulations. Together with the apparent constancy of $z_q$, and thus constant $v(z > 0)$, in natural saltation, this puzzling discrepancy suggests that wind tunnel experiments and models neglect or misrepresent one or more processes that enhance particle trajectories and/or suppress the growth of $z_q$ with $u_*$. One candidate process is electrification, which could provide a downward force that increases with $u_*$ *(54)*. This could explain the constancy of $z_q$ with $u_*$, but it does not explain the underestimation of the saltation layer height. A second candidate process is mid-air interparticle collisions, but these have a dispersive effect expected to cause an enhancement of both $z_q$ and $Q$ that increases with $u_*$ *(55, 56)*, which contrasts with our observations. A third candidate process is modification of saltation transport mechanics by atmospheric turbulence *(45, 49)*. Turbulence structures span a much wider range of length and time scales in the field than is represented in wind tunnels *(48)*, though it is not yet clear how large-scale turbulence structures affect saltation *(57)*. It is also possible that the small boundary layer depth comparable to saltation height in wind tunnels *(29)* affects the trajectories of energetic particles amidst turbulent winds. Given our poor understanding of how these candidate processes affect saltation trajectories, they require further exploration.

By confirming constant saltation heights and a linear flux law, our findings also provide insight into how saltation mechanics and flux scaling might vary with atmospheric properties on Mars and other planetary bodies. On Earth, constant saltation heights can be explained by splash-dominated particle entrainment, which balances numbers of impacting and ejecting particles by maintaining constant near-surface particle speed $v_0$ *(48, 51)*. In contrast, fluvial bedload, in which particle entrainment is dominated by fluid lifting, accommodates linear scaling of particle speed with shear velocity and therefore nonlinear 3/2 flux scaling *(58)*. Fluvial bedload and aeolian saltation are distinguished by a large difference in particle-fluid density ratio, $s = \rho_p/\rho_f$, which appears to control the relative importance of splash versus fluid entrainment of particles *(59, 60)*. We therefore expect *s* also to determine the occurrence of linear versus nonlinear flux scaling under transport-limited conditions *(61)*. When *s* is between the Earth aeolian ($s \approx 2000$) and fluvial cases ($s \approx 2.65$), as in aeolian transport on Venus ($s \approx 40$) and Titan ($s \approx 190$) *(60)*, we expect a scaling of the flux with stress that is intermediate between the linear (aeolian) and nonlinear (fluvial) cases. However, on planetary surfaces where $s \geq 2000$, we predict that particle speeds and saltation layer heights remain constant with shear velocity. Furthermore, the dominance of splash entrainment in these cases suggests a clear separation between the 'fluid threshold' stress for initiating saltation and a lower 'impact threshold' for sustaining it *(62, 63)*.

We thus conclude that saltation flux scales linearly with fluid stress on planetary bodies with atmospheres equally or more dilute than Earth's. A linear saltation flux law therefore should apply to studies of ripples, dunes, dust aerosol emission, and other aeolian processes on Earth ($s \approx 2000$), Mars ($s \approx 2.5 \times 10^5$), Neptune's moon Triton ($s \approx 10^7$), Jupiter's moon Io ($s \approx 10^{12}$), comets (e.g., $s \approx 10^{12}$) *(60)*, and possibly even Pluto ($s \approx 10^7$) *(64)*. Given that the study of aeolian processes on Earth and these planetary bodies has long been dominated by the use of nonlinear saltation flux laws *(13, 19–22, 65)*, our finding that saltation flux instead scales linearly with wind stress has implications for our understanding of a wide range of processes across the solar system.

## Materials and Methods
*Experimental design*
We obtained coupled field measurements of wind velocity and aeolian sediment transport profiles at three sites: Jericoacoara, Ceara, Brazil; Rancho Guadalupe, California, United States; and Oceano, California, United States, providing respectively 3, 2, and 12 measurement days of active saltation *(43)*. All measurement sites were relatively flat with ≈100–300 m of open sand upwind and ≈0.6–1.0 km of upwind fetch to the shoreline. We chose sites that represent a range of soil conditions and size distributions. Surface sediment samples were collected daily, and the grain size distributions for these samples were determined using a Retsch Camsizer particle size analyzer *(43)*. We calculated median surface particle diameters $d_{50}$ (Table 1) from the average of all sample distributions at each site *(43)*.

*Wind calculations*
We determined wind shear stress from sonic anemometer observations (25 Hz at Jericoacoara and Rancho Guadalupe; 50 Hz at Oceano) at about half a meter above the surface. As in Kok et al. *(66)*, we subdivided the wind data time series into continuous 30-minute intervals: long enough to capture the largest turbulence structures but short enough to resolve meteorological variability *(39)*. For each of these intervals, we rotated the coordinate system according to the procedure in van Boxel et al. *(39)*, so that interval mean lateral ($u$) and vertical ($w$) winds were 0. Over each of these time intervals, we computed the time-averaged shear stress by the Reynolds stress method as:
$$\tau = -\rho_f \overline{u'w'} = -\rho_f \overline{(u - \bar{u})(w - \bar{w})}, \qquad (6)$$
where $u'$ and $w'$ ($\bar{u}$ and $\bar{w}$) are the fluctuating (mean) components of the streamwise and vertical wind, respectively, and the overbar refers to the ensemble average over the entire 30-minute interval. Based on a typical temperature of 30°C at Jericoacoara and 15°C at Rancho Guadalupe and Oceano measured during the deployments, we used in all calculations air density values of $\rho_f = 1.16$ kg/m³ and 1.22 kg/m³ for these respective sites corresponding to air pressure at sea level on Earth. In all analyzed 30-minute intervals, winds were aligned with the prevailing wind direction (wind angle $|\theta| = |\tan^{-1}(\bar{v}/\bar{u})| \leq 20°$) and conditions were roughly neutrally stable (stability parameter $|z/L| \leq 0.15$) *(43)*.

*Saltation flux calculations*
At each site, vertical profiles of horizontal saltation flux $q(z)$ were determined from Wenglor laser particle counter *(67)* measurements (25 Hz) at multiple (3–9) heights ($z = 0.02$–0.47 m). We calibrated Wenglor counts to 1-hour absolute mass fluxes from BSNE saltation trap *(50)* measurements at multiple (4–8) heights ($z = 0.05$–0.70 m) fitted to the exponential profile in equation 2 *(43)*. We chose this method because the Wenglors provided much higher temporal resolution, but counting sensitivities among Wenglors varied substantially in space and time due to scratching and dust buildup on lenses and other unexplained factors. From the calibrated Wenglor fluxes, we subsampled 30-minute averaged $q(z)$ profiles. Based on these profiles (Fig. 6 in Martin et al. *(43)*), we again applied the exponential fit (equation 2) to calculate characteristic saltation layer height $z_q$ (Fig. 7 in Martin et al. *(43)*) and total saltation flux $Q$ (Fig. 8 in Martin et al. *(43)*). To account for variability in measured fluxes and saltation layer heights, we combined 30-minute values into bins defined by ranges of shear stress. We then computed the mean and uncertainty of $z_q$ and $Q$ for each bin. When generating binned values for $z_q$, we included only 30-minute values with detected transport, as $z_q$ is undefined when transport is not occurring (see supplementary text).

*Derivation of linear flux law*
Based on our finding that $z_q$ is constant, we derived a scaling relationship for the total saltation flux. The saltation flux is driven by that portion of the wind stress that is not dissipated through friction at the soil surface. Measurements and theory indicate that the fluid stress dissipated at the bed during steady state saltation approximately equals the impact threshold stress $\tau_{it}$ required to sustain saltation *(16, 18, 29, 30)*. For $\tau > \tau_{it}$, the 'excess' stress available to do work moving sediment is then:
$$\tau_{ex} = \tau - \tau_{it}. \qquad (7)$$
In equilibrium, the momentum input $\tau_{ex}$ is balanced by particle momentum dissipation due to inelastic saltator collisions with the soil bed $\tau_p$, i.e.,
$$\tau_{ex} = \tau_p = MV(1-e), \qquad (8)$$
where $M$ [gm$^{-2}$] is the mass collision rate per unit bed area and $e$ is the (dimensionless) bulk restitution coefficient of saltator impacts. $M$ is related to the vertically integrated particle concentration $\Phi$ by:
$$\Phi = Mt_{hop}, \qquad (9)$$
where $t_{hop}$ is the mean saltator hop time. To first order, we assume particle trajectories to be ballistic *(48, 51)*, for which $t_{hop}$ is related to the mean hop height $z_{hop}$ as:
$$t_{hop} = \sqrt{\frac{8z_{hop}}{g}}. \qquad (10)$$
Assuming that the saltation layer *e*-folding height $z_q$ depends on the mean particle trajectory height *(17, 18, 23)*, we have:
$$z_{hop} = C_z z_q, \qquad (11)$$
where $C_z$ is a (dimensionless) constant of order 1. Combining this with equations 1, 8, and 9 yields an expression for the saltation mass flux:
$$Q = C_t \frac{\sqrt{z_q/g}}{1-e} \tau_{ex}. \qquad (12)$$
Since our field measurements show that $z_q \sim d_{50}$, and since we further have that $u_{*,it} \sim \sqrt{gd_{50}}$ *(17)*, equation 12 can be simplified to
$$Q = C_Q \frac{u_{*,it}}{g} \tau_{ex}, \qquad (13)$$
where $C_Q$ is a (dimensionless) scaling parameter. Equation 12 more explicitly includes the physical parameters ($z_q$, $e$, $\tau_{it}$) that determine the saltation flux law, whereas equation 13 is a simpler expression facilitating comparisons among studies.

*Statistical Analysis*

We compared linear and nonlinear 3/2 flux laws by fitting binned stress and flux data to these two scalings. When performing fits, we included only $\tau$ bins for which transport was detected at least 10% of the time, because a flux law only applies during transport. We determined fits to the linear flux law ($Q = C(\tau - \tau_{it})$) through linear regression, and we determined fits to the nonlinear 3/2 flux law ($Q = Cu_*(\tau - \tau_{it})$) by finding the values of the parameters ($C$ and $\tau_{it}$) that minimized the value of $\chi^2$, the mean-square difference between observations and predictions (see supplementary text and Table S1). Based on our finding of a linear scaling, we used $\tau_{it}$ from the linear fit to compute excess stress $\tau_{ex}$ (equation 3). We also used this $\tau_{it}$ to estimate the dimensionless flux scaling parameter $\hat{C}_Q$ (equation 4).

## Supplementary Information

Supplementary text
Fig. S1. Standard deviation of binned values for saltation flux versus binned saltation flux.
Fig. S2. Binned saltation flux versus binned shear stress.
Table S1. Saltation flux law fit values for the three field sites.

## Acknowledgments:

**General**: Oceano Dunes State Vehicular Recreation Area, Rancho Guadalupe Dunes Preserve, and Jericoacoara National Park provided essential site access and support. Jericoacoara fieldwork is registered with the Brazilian Ministry of the Environment (#46254-1 to J. Ellis). We thank Marcelo Chamecki for advice on treatment of wind data, Chris Hugenholtz and Tom Barchyn for equipment help, Doug Jerolmack for lab access for grain-size analysis, Kate Ledger for writing guidance, and Jean Ellis, Paulo Sousa, Peter Li, Francis Turney, Arkayan Samaddar, and Livia Freire for field assistance.

**Funding:** U.S. National Science Foundation (NSF) Postdoctoral Fellowship EAR-1249918 to R.L.M., NSF grant AGS-1358621 to J.F.K., and ARO grant W911NF-15-1-0417 to J.F.K. supported this research.

**Author contributions:** R.L.M. led and J.F.K. closely advised all field data collection, data analysis, and writing.

**Competing interests:** The authors declare no competing financial interests.

**Data and materials availability:** Unbinned and binned analysis values may be accessed through the Zenodo data repository at http://doi.org/10.5281/zenodo.291798 and http://doi.org/10.5281/zenodo.291799, respectively.

# Supplementary Information

## Overview
In this supplementary document, we present detailed descriptions of data binning procedures, sources of literature data, and data fitting. This document elaborates on primary data processing operations described in Martin et al. *(43)*. We conclude with a detailed list of all variables used in the main text and supplementary information, two supplementary figures, and one supplementary table.

## Data binning and uncertainty estimation
In this section, we describe our procedures for aggregating calculations from individual 30-minute time intervals into bins covering ranges of shear stress $\tau$ for each field site. By sorting data into bins, we were able to obtain statistically representative values for saltation height $z_q$ and saltation flux $Q$ and quantify the associated uncertainties for these values. Binning also provided an objective basis for conditioning analyses on specific ranges of transport frequency. To distinguish unbinned versus binned values in this section, we apply the tilde symbol (~) to unbinned 30-minute values and their uncertainties, whereas '$i$' subscripts are used to denote each binned value and its uncertainty. All unbinned 30-minute values were obtained in Martin et al. *(43)*.

*Creation of shear stress bins*
For each site, we categorized each 30-minute value into a bin based on the 30-minute shear stress $\tilde{\tau}$ value during the associated time interval. Therefore, each bin $i$ was defined by a range of shear stress values, $[\tau_{min,i}, \tau_{max,i}]$. Ideally, such bins would have been uniformly constructed such that $\tau_{max,i} - \tau_{min,i}$ was the same for all bins. However, due to natural fluctuations in the wind, the $\tilde{\tau}$ were not equally spaced across the full range of possible shear stresses at each site. To balance the priorities of including sufficient data points in each bin while avoiding creation of bins spanning excessively wide ranges of $\tau$, we developed a binning protocol for each site as follows:

(1) Sort all 30-minute data points in order of increasing $\tilde{\tau}$. For the first bin, set $\tau_{min,1}$ as the minimum of all the $\tilde{\tau}$ for the site.
(2) Add points to this bin in order of increasing $\tilde{\tau}$ until the bin is considered full by the following criterion:
   a. $\tau_{max,i} - \tau_{min,i} \geq 0.01$ Pa, AND
   b. There are at least 3 points in the bin OR $\tau_{max,i} - \tau_{min,i} > 0.025$ Pa.
(3) Once the current bin $i$ is full, create a new bin $i$+1 with $\tau_{min,i+1}$ equal to the next value of $\tilde{\tau}$ from the ordered list. Then repeat step 2.

Condition 2a was created to ensure that each bin covers a reasonably wide range of $\tau$ (comparable to typical relative errors in $\tau$), while condition 2b was created to balance the priorities of including sufficient data points in bins without creating excessively wide bins. Though our selection of binning criteria was necessarily arbitrary, we found that reasonable changes in our binning protocol would not have qualitatively affected any of our conclusions.

*Estimation of binned values and their uncertainties*

Let $\tilde{a}_j$ and $\tilde{\sigma}_{a_j}$ denote unbinned 30-minute values and their uncertainties for variable $a$ within a certain $\tau$ bin $i$. We computed each binned value $a_i$ as an arithmetic average:

$$a_i = \sum_j \tilde{a}_j / N_i, \qquad (S1)$$

where $N_i$ is the number of values in the bin. We estimated the binned value uncertainty $\sigma_{a_i}$ by two separate methods: (1) propagation of the random errors of the constituent data points, and (2) the standard error of the $N_i$ values of $\tilde{a}_j$. We computed the uncertainty associated with random errors $\sigma_{\mu,a_i}$ (equation 4.19 in Bevington and Robinson *(68)*) as:

$$\sigma_{\mu,a_i} = \sqrt{\frac{1}{\sum_j (1/\tilde{\sigma}_{a_j}^2)}}. \qquad (S2)$$

This calculation represents the uncertainty associated with measurement errors that we can quantify and that are discussed in Martin et al. *(43)*. We computed the bin standard error $\sigma_{SE,a_i}$ as (equation 4.14 in Bevington and Robinson *(68)*):

$$\sigma_{SE,a_i} = \frac{\sqrt{\frac{1}{N_i}\sum_j (\tilde{a}_j - a_i)^2}}{\sqrt{N_i}}. \qquad (S3)$$

This calculation represents the uncertainty associated with differences in measured values due to all sources of error, including those that we were not able to quantify, such as changes in soil conditions. We then estimated the overall uncertainty in $a_i$ as:

$$\sigma_{a_i} = \max(\sigma_{SE,a_i}, \sigma_{\mu,a_i}). \qquad (S4)$$

Conservatively taking the bin uncertainty as the maximum of these two methods helped to ensure that a realistic error was assigned to a binned value, such as in cases where the bin contained few constituent values, or when unquantified systematic errors contributed a substantial fraction of the total uncertainty.

Though the above binning procedure was meant to provide sufficient values in each bin to compute a standard error, sometimes a bin contained insufficient values for computation of standard error due to condition (1) in the previous subsection. We denote these bins with insufficient values (those with fewer than 3 data points) as 'sparse' bins. We estimated 'modified' standard errors $\sigma'_{SE,a_i}$ for sparse bins based on 'comparable' bins at the same site containing 3 or more data points. For each comparable bin $i$, we computed the standard deviation of the values in the bin $SD_{a_i}$, then we computed the median of these values $SD_{med,a}$. The calculation for $SD_{med,a}$ is illustrated in Fig. S1 for saltation flux $Q$. Based on this typical standard deviation for binned values, we then estimated modified standard errors for the sparse bins as:

$$\sigma'_{SE,a_i} = \frac{SD_{med,a}}{\sqrt{N_i}}, \qquad (S5)$$

where $N_i$ is the number of data points in the bin. For the sparse bins, values of $\sigma'_{SE,a_i}$ were then used instead of $\sigma_{SE,a_i}$ for calculation of $\sigma_{a_i}$ in equation S4. For saltation flux, a few of the calculated modified standard errors $\sigma'_{SE,Q_i}$ unrealistically exceeded the associated flux values $Q_i$. In these cases, we adjusted the values so that $\sigma'_{SE,Q_i} = Q_i$.

We applied the above binning procedure to 30-minute values for $\tilde{\tau}, \tilde{u}_*, \tilde{Q}$. This yielded binned values $\tau_i$, $u_{*,i}$, and $Q_i$, with associated bin uncertainties $\sigma_{\tau_i}$, $\sigma_{u_{*,i}}$, and $\sigma_{Q_i}$. For computation of binned saltation layer heights $z_{q,i}$ and associated uncertainties $\sigma_{z_{q,i}}$, we included only those 30-minute values $\tilde{z}_q$ with corresponding 30-minute saltation fluxes that were nonzero, i.e., $\tilde{Q} > 0$. We applied this restriction because $\tilde{z}_q$ is undefined for $\tilde{Q} = 0$. The binning procedure is illustrated by comparing Fig. 7 in Martin et al. *(43)*, which shows the unbinned $\tilde{z}_q$ versus $\tilde{u}_*$, to Fig. 1A (main text), which shows binned $z_q$ versus $u_*$.

## Analysis of literature data

In this section, we describe analysis of the three literature data sets for saltation layer heights: Greeley et al. (1996), Namikas (2003), and Farrell et al. (2012). In Fig. 1 of the main text, we compared saltation layer heights $z_q$ computed from these literature data to the values of $z_q$ calculated at our own field sites.

*Greeley et al. (1996)*
We obtained data from the flux profiles shown in Fig. 11 of Greeley et al. *(40)*. We estimated $z_q$ and corresponding uncertainty $\sigma_{z_q}$ for each partial (height-specific) flux profile based on an exponential fit (equation 11 in Martin et al. *(43)*). We obtained shear velocities $u_*$ for corresponding runs directly from Table 1 in the main text. We chose $d_{50} = 0.23$ mm based on the "modal distribution" value stated in the paper. Though exact coordinates for the field site were not provided, the site described in the paper appears to be located close to our Oceano site.

*Namikas (2003)*
We obtained data from the flux profiles shown in Fig. 6 of Namikas *(41)*. We estimated $z_q$ and corresponding uncertainty $\sigma_{z_q}$ for each partial flux profile based on an exponential fit (equation 11 in Martin et al. *(43)*). We obtained shear velocities $u_*$ for corresponding runs directly from Table 2 in the Namikas paper. We chose $d_{50} = 0.25$ mm based on the "averaged" value stated in the paper. Though exact coordinates for the field site were not provided, the site described in the paper appears to be close to our Oceano site.

*Farrell et al. (2012)*
We obtained shear velocities and flux profiles based on values listed in Table 1 of Farrell et al. *(42)*. For the analysis, we considered all runs at the "Cow Splat Flat Fine (CSFF)" site that included a value for $u_*$, except for Run 1, for which the trap duration was significantly shorter than the other runs. Based on the site coordinates listed in the paper, CSFF was located within a few hundred meters of our Jericoacoara site. We ignored data from the "Cow Splat Flat Coarse (CSFC)" and "BEACH" sites listed in the paper, because these had different soil characteristics from CSFF but contained few data points. For each listed run at CSFF, we estimated partial flux $q_i$ for each trap based on its collected mass, trap width, trap height, and duration, according to the same procedures described for our BSNE partial flux estimation (equation 24 in Martin et al. *(43)*). Table 1 in Farrell et al. provided only the bottom and top heights of the stacked traps, so we performed the profile fitting iteratively to estimate trap heights $z_i$ and saltation layer height $z_q$ as in equation 29 in Martin et al. *(43)*. Since Farrell et al. did not report the particle size distribution at the surface, we could not estimate $d_{50}$ for our analysis of non-dimensional saltation heights.

## Data fitting and derivation of parameters
In this section, we describe methods for fitting to binned wind and saltation data to characterize saltation height versus shear velocity and saltation flux versus shear stress trends. These fits correspond to the values listed in Table 1 of the main text.

*Fitting saltation height versus shear velocity*
In this section, we describe calculations of the mean saltation height and appraisal of its linear trend versus shear velocity. For calculations at our field sites, $z_{q,i}$ and $u_{*,i}$ refer to the individual binned values for each site. For the literature sites, $z_{q,i}$ and $u_{*,i}$ refer directly to the values from individual profile fits (i.e., no binning).

For each site, we performed a linear fit to saltation height $z_{q,i}$ versus shear velocity $u_{*,i}$, as:
$$z_{q,fit,i} = a + bu_{*,i}, \tag{S6}$$
where $z_{q,fit,i}$ are the predicted values for the linear fit. Intercept $a$, slope $b$, and associated uncertainties in these parameter fits ($\sigma_a$ and $\sigma_b$) were calculated by the linear fitting procedure described in Martin et al. *(43)*. Values for $b$ are shown in Fig. 1B (main text).

We calculated the mean saltation layer height for each site (Table 1 of main text) as:
$$\langle z_q \rangle = \frac{\sum_i z_{q,i}}{N}, \tag{S7}$$
where $N$ is the number of bins for each site. The corresponding uncertainty in the mean saltation layer height was then calculated based on the standard deviation:
$$\sigma_{\langle z_q \rangle} = \sqrt{\frac{1}{N}\sum_i (z_{q,i} - \langle z_q \rangle)^2}. \tag{S8}$$
We used the standard deviation rather than the standard error here because differences in individual $z_{q,i}$ reflected actual variability rather than measurement uncertainty.

The mean dimensionless saltation layer height is simply the ratio of the mean saltation layer height and the median surface particle diameter, $\langle z_q \rangle / d_{50}$. We calculated the associated uncertainty in dimensionless saltation layer height as:
$$\sigma_{\langle z_q \rangle / d_{50}} = \frac{1}{d_{50}} \sqrt{\sigma_{\langle z_q \rangle}^2 + \sigma_{d_{50}}^2 \left(\langle z_q \rangle / d_{50}\right)^2}. \tag{S9}$$
To obtain equation S9, we applied the error propagation equation for uncorrelated variables (equation 3.14 in Bevington and Robinson *(68)*),
$$\sigma_y = \sqrt{\sigma_a^2 \left(\frac{\partial y}{\partial a}\right)^2 + \sigma_b^2 \left(\frac{\partial y}{\partial b}\right)^2 + \cdots}, \tag{S10}$$
where the variable $y$ is a function of quantities $a, b, \ldots$ with uncorrelated uncertainties $\sigma_a, \sigma_b, \ldots$.

*Fitting saltation flux versus shear stress*
In this section, we describe our methods for performing fits to both the linear and the nonlinear saltation flux law. These fits considered binned values of shear stress $\tau_i$ and total saltation flux $Q_i$ at each site. Because the saltation flux law only applies when shear stress is above a threshold value $\tau_{it}$, we removed certain low stress bins from the analysis for each site. However, since we lacked knowledge of $\tau_{it}$ prior to performing the fit, we required independent criteria for selecting the binned values to include in the fit. For this criterion, we chose a minimum transport frequency, below which transport is very intermittent and equilibrium saltation transport conditions may never occur. In particular, we limited analysis to bins $i$ for which saltation was detected for at least 10% of 1-second increments within the bin. The resulting binned values $Q_i$ and $\tau_i$ for the fit, subject to the 10% limit, are shown in Fig. S2.

Based on equation 3 (main text), we estimated parameters $C$ and $\tau_{it}$ for a linear flux law at each site based on a linear fit to
$$Q_{fit,i} = C(\tau_i - \tau_{it}). \tag{S11}$$
We performed the fit to equation S11 based on the linear fitting procedure described in Martin et al. *(43)*. The resulting linear fits are shown in Fig. S2. The resulting values of $C$ and $\tau_{it}$ for each site and their associated uncertainties ($\sigma_C$ and $\sigma_{\tau_{it}}$) are listed in Table S1.

We evaluated the quality of these fits by computing the reduced chi-square $\chi_\nu^2$, which expresses the difference between the best fit and the measured values, normalized by the number of degrees of freedom $\nu$ (number of data points minus number of fitting parameters, which is 2 for the linear fit):
$$\chi_\nu^2 = \frac{1}{\nu}\sum_i \frac{(Q_i - Q_{fit,i})^2}{\sigma_{Q_i,total}^2}, \tag{S12}$$
where $\sigma_{Q_i,total}$ is the combined uncertainty for the binned flux values $\sigma_{Q_i}$ and the propagated uncertainty in binned shear stress values $\sigma_{\tau_i}$. The reduced chi-square values for the linear fits are listed in Table S1. $\chi_\nu^2$ describes the extent to which the fitted function explains the measured data and its uncertainty. A value of $\chi_\nu^2 \approx 1$ indicates that the model reasonably describes the observational data within the uncertainties, whereas a value of $\chi_\nu^2 \gg 1$ indicates either that the model does not capture all of the variance of the data, or that the uncertainty in the measurements is underestimated *(68)*.

To test the validity of the linear flux law, we compared the quality of linear fits to alternative 3/2 fits. The nonlinear 3/2 fit equation is:
$$Q_{fit,i} = Cu_{*,i}(\tau_i - \tau_{it}). \tag{S13}$$
We numerically determined the values of $C$ and $\tau_{it}$ that minimize $\chi_\nu^2$ (equation S12). In determining each $\sigma_{Q_i,total}$ for the $\chi_\nu^2$ calculation, we computed uncertainty both in the actual measured flux $\sigma_{Q_i}$ and the flux uncertainty propagated from uncertainty in the shear stress for the combination of fit values:
$$\sigma_{Q,\tau_i} = \sigma_{u_{*,i}} C |3\tau_i - \tau_{it}|, \tag{S14}$$
where equation S14 was computed using the error propagation formula (equation S10). For computation of $\chi_\nu^2$, we calculated the total flux uncertainty for each point as:
$$\sigma_{Q_i,total} = \sqrt{\sigma_{Q_i}^2 + \sigma_{Q,\tau_i}^2}. \tag{S15}$$
Nonlinear 3/2 flux fits are compared to the linear fits in Fig. S2, and the associated $\chi_\nu^2$ values are listed in Table S1. For the nonlinear fits, uncertainties on fitting parameters ($\sigma_C$ and $\sigma_{\tau_{it}}$) were determined as the ranges of $C$ and $\tau_{it}$ for which $\chi^2 \leq \min(\chi^2) + 1$, where $\chi^2$ is the non-reduced chi-square value *(68)*.

At Jericoacoara and Oceano, the values for $\chi_\nu^2$ were substantially smaller for the linear fit than for the nonlinear 3/2 fit, indicating the superiority of the linear flux law. At Rancho Guadalupe, $\chi_\nu^2$ values were similar for linear and nonlinear 3/2 fits, indicating that neither fit was preferable. At Jericoacoara and Oceano, $\chi_\nu^2 > 1$ indicates some underestimation of the saltation flux uncertainties. At Rancho Guadalupe, $\chi_\nu^2 < 1$ indicates some overestimation of the saltation flux uncertainties. These differences might reflect that data at these sites were insufficient to fully characterize the variability in saltation flux.

*Calculation of excess stress*
We calculated the excess stress as:
$$\tau_{ex,i} = \tau_i - \tau_{it}, \tag{S16}$$
where $\tau$ is the binned shear stress and $\tau_{it}$ is the impact threshold from the linear fit (equation S11, Table S1). $\tau_{ex,i}$ is defined only for $\tau_i > \tau_{it}$, so for subsequent calculations in this section, we included only binned values $i$ for which all of the individual $\tilde{\tau}$ in the bin exceed $2\sigma_{\tau_{it}}$, i.e., outside the 95% confidence range for $\tau_{it}$.

The uncertainty in excess stress $\sigma_{\tau_{ex,i}}$ depended both on uncertainty in binned values of shear stress $\sigma_{\tau_i}$ and uncertainty in the threshold stress $\sigma_{\tau_{it}}$ through error propagation (equation S10):
$$\sigma_{\tau_{ex,i}} = \sqrt{\sigma_{\tau_i}^2 + \sigma_{\tau_{it}}^2}. \tag{S17}$$
We also calculated the dimensionless stress $\tau_i/\tau_{it}$, whose associated uncertainty was determined through error propagation (equation S10) as:
$$\sigma_{\tau_i/\tau_{it}} = \frac{1}{\tau_{it}}\sqrt{\sigma_{\tau_i}^2 + \sigma_{\tau_{it}}^2\left(\frac{\tau_i}{\tau_{it}}\right)^2}. \tag{S18}$$
Based on the linear best fit shear stress threshold $\tau_{it}$, we estimated the associated shear velocity threshold as $u_{*it} = \sqrt{\tau_{it}/\rho_f}$. We then calculated the corresponding uncertainty in shear velocity threshold by error propagation (equation S10):
$$\sigma_{u_{*it}} = \frac{\sigma_{\tau_{it}}}{2\sqrt{\rho_f \tau_{it}}}. \tag{S19}$$
The values of shear velocity threshold and their associated uncertainties are listed in Table 1 (main text).

*Calculation of dimensionless flux scaling parameter*
We estimated the dimensionless flux scaling parameter for stress bin $i$ by equation 4 (main text):
$$\hat{C}_{Q,i} = \frac{g}{u_{*,it}} \frac{Q_i}{\tau_{ex,i}}. \tag{S20}$$
We computed each $\hat{C}_{Q,i}$ based on the fitted value for $u_{*,it}$ at each site and the binned values $Q_i$ and $\tau_{ex,i}$ at each site. Based on error propagation (equation S10), the contributions to the uncertainty in $\hat{C}_{Q,i}$ include the uncertainty in the flux $\sigma_{Q_i}$, uncertainty in the excess stress $\sigma_{\tau_{ex,i}}$, and uncertainty in the impact threshold $\sigma_{u_{*it}}$:
$$\sigma_{\hat{C}_{Q,i}} = \hat{C}_{Q,i}\sqrt{\left(\frac{\sigma_{Q_i}}{Q_i}\right)^2 + \left(\frac{\sigma_{\tau_{ex,i}}}{\tau_{ex,i}}\right)^2 + \left(\frac{\sigma_{u_{*,it}}}{u_{*,it}}\right)^2}. \tag{S21}$$
We calculated the mean value for $\hat{C}_Q$ for each site as:

$$\langle \hat{C}_Q \rangle = \frac{\sum_i \hat{C}_{Q,i}}{N}, \tag{S22}$$

where $N$ is the number of bins included in the mean for each site. Corresponding uncertainty in $\langle \hat{C}_Q \rangle$ was then calculated based on the standard deviation:

$$\sigma_{\langle \hat{C}_Q \rangle} = \sqrt{\frac{1}{N}\sum_i(\hat{C}_{Q,i} - \langle \hat{C}_Q \rangle)^2}. \tag{S23}$$

As with saltation layer height (equation S7), we used the standard deviation rather than the standard error here because differences in individual $\hat{C}_{Q,i}$ reflected actual variability rather than measurement uncertainty. Based on equation 3 (main text) we estimated the parameter $C_Q$ in the linear flux law and its associated uncertainty $\sigma_{C_Q}$ directly from $\langle \hat{C}_Q \rangle$ and $\sigma_{\langle \hat{C}_Q \rangle}$, respectively. The resulting values of $C_Q$ and $\sigma_{C_Q}$ are given in Table 1 (main text).

An alternative version of the dimensionless flux scaling parameter is $C_t$ (equation 12 in main text), which we estimate for stress bin $i$ as:

$$\hat{C}_{t,i} = (1-e)\sqrt{\frac{g}{\langle z_q \rangle}\frac{Q_i}{\tau_{ex,i}}}. \tag{S24}$$

Because the restitution coefficient $e$ is unknown, we instead estimated the parameter $\hat{C}_{t,i}/(1-e)$. Based on error propagation (equation S10) the uncertainty in $\hat{C}_{t,i}$ includes contributions of uncertainty in the flux $\sigma_{Q_i}$, excess stress $\sigma_{\tau_{ex,i}}$, and mean saltation height $\sigma_{\langle z_q \rangle}$:

$$\frac{\sigma_{\hat{C}_{t,i}}}{(1-e)} = \hat{Q}_i\sqrt{\left(\frac{\sigma_{Q_i}}{Q_i}\right)^2 + \left(\frac{\sigma_{\tau_{ex,i}}}{\tau_{ex,i}}\right)^2 + \left(\frac{\sigma_{\langle z_q \rangle}}{2\langle z_q \rangle}\right)^2}. \tag{S25}$$

For each site, we computed the mean value of $\hat{C}_t/(1-e)$ and the uncertainty of this mean, in the same manner as we did for $\hat{C}_Q$ (equations S22 and S23). We then used these mean values to estimate the parameter $C_t/(1-e)$ in equation 12 (main text) and its associated uncertainty $\sigma_{C_t}/(1-e)$ for each site (Jericoacoara: $2.5 \pm 0.3$, Rancho Guadalupe: $1.7 \pm 0.1$, Oceano: $2.2 \pm 0.4$).

# List of variables
Below, we list all variables described in the main text and supplementary text. Typical units for variables are given in parentheses.

*Primary variables and physical quantities*
$d_{50}$, median diameter of surface particles by volume (mm)
$e$, bulk restitution coefficient
$g$, gravitational acceleration (m s$^{-2}$)
$\rho_f$, air density (kg m$^{-3}$)
$\rho_p$, particle density (kg m$^{-3}$)
$s = \rho_p/\rho_f$, particle-fluid density ratio
$t_{hop}$, typical hop time for saltating particles (s)
$z_{hop}$, typical maximum hop height for saltating particles (m)
$z_q$, e-folding saltation layer height (m)
$M$, mass collision rate per unit bed area (kg m$^{-2}$ s$^{-1}$)
$Q$, vertically-integrated saltation sand flux (g m$^{-1}$ s$^{-1}$)
$\Phi$, vertically-integrated saltation layer mass concentration (kg m$^{-2}$)
$V$, mean horizontal particle speed (m s$^{-1}$)
$v_0$, mean near-surface horizontal particle speed (m s$^{-1}$)
$v(z)$, mean horizontal particle speed as a function of height (m s$^{-1}$)
$\tau$, wind shear stress (Pa = kg m$^{-1}$ s$^{-2}$)
$\tau_{ex}$, excess shear stress (Pa)
$\tau_{it}$, impact threshold shear stress (Pa)
$\tau/\tau_{it}$, dimensionless shear stress
$\tau_p$, particle momentum dissipation rate (Pa)
$u_*$, wind shear velocity (m s$^{-1}$)
$u_{*,it}$, impact threshold wind shear velocity (m s$^{-1}$)
$\kappa$, von Karman parameter ($\approx$0.4)
$f$, scaling exponent for $Q$ versus $\tau$
$c$, scaling exponent for $\Phi$ versus $\tau$
$r$, scaling exponent for $V$ versus $\tau$

*Wind variables*
$u$, streamwise wind velocity (m s$^{-1}$)
$u'$, fluctuating component of streamwise wind (m s$^{-1}$)
$\bar{u}$, mean streamwise wind velocity (m s$^{-1}$)
$\bar{v}$, mean spanwise wind velocity (m s$^{-1}$)
$w$, vertical wind velocity (m s$^{-1}$)
$w'$, fluctuating component of vertical wind (m s$^{-1}$)
$\bar{w}$, mean vertical wind velocity (m s$^{-1}$)
$T$, air temperature (K)
$z/L$, stability parameter
$\theta$, wind angle
$\tilde{\tau}$, 30-minute shear stress (Pa)
$\tilde{\sigma}_\tau$, 30-minute shear stress uncertainty (Pa)

$\tau_i$, wind shear stress for stress bin $i$ (Pa)
$\sigma_{\tau_i}$, uncertainty in wind shear stress for stress bin $i$ (Pa)
$\tau_{min,i}$, minimum wind shear stress for stress bin $i$ (Pa)
$\tau_{max,i}$, maximum wind shear stress for stress bin $i$ (Pa)
$\tau_{ex,i}$, excess stress for stress bin $i$ (Pa)
$\sigma_{\tau_{ex,i}}$, uncertainty in excess stress for stress bin $i$ (Pa)
$\tau_i/\tau_{it}$, dimensionless shear stress for stress bin $i$
$\sigma_{\tau_i/\tau_{it}}$, uncertainty in dimensionless shear stress for stress bin $i$
$\tilde{u}_*$, 30-minute shear velocity (m s$^{-1}$)
$\tilde{\sigma}_{u_*}$, 30-minute shear velocity uncertainty (m s$^{-1}$)
$u_{*,i}$, shear velocity for stress bin $i$ (m s$^{-1}$)
$\sigma_{u_{*,i}}$, uncertainty in shear velocity for stress bin $i$ (m s$^{-1}$)

*Flux variables*
$q_0$, saltation profile scaling parameter (g m$^{-2}$ s$^{-1}$)
$\sigma_{q_0}$, uncertainty in saltation profile scaling parameter (g m$^{-2}$ s$^{-1}$)
$\tilde{Q}$, 30-minute total saltation flux (g m$^{-1}$ s$^{-1}$)
$\tilde{\sigma}_Q$, 30-minute total saltation flux uncertainty (g m$^{-1}$ s$^{-1}$)
$Q_i$, total saltation flux for stress bin $i$ (g m$^{-1}$ s$^{-1}$)
$\sigma_{Q_i}$, uncertainty in total saltation flux for stress bin $i$ (g m$^{-1}$ s$^{-1}$)
$\sigma_{Q,\tau_i}$, uncertainty in saltation flux for stress bin $i$ due to uncertainty in shear stress (g m$^{-1}$ s$^{-1}$)
$\sigma_{Q_i,total}$, total saltation flux uncertainty (including propagated stress uncertainty) (g m$^{-1}$ s$^{-1}$)
$Q_{fit,i}$, fitted value for total saltation flux for stress bin $i$ (g m$^{-1}$ s$^{-1}$)
$SD_{Q_i}$, standard deviation for binned saltation flux $Q_i$ values (g m$^{-1}$ s$^{-1}$)
$\sigma'_{SE,Q_i}$, modified standard error in total saltation flux for stress bin $i$ (g m$^{-1}$ s$^{-1}$)
$z_q$, characteristic *e*-folding saltation layer height (m)
$\sigma_{z_q}$, uncertainty in characteristic *e*-folding saltation layer height (m)
$\tilde{z}_q$, 30-minute characteristic *e*-folding saltation layer height (m)
$\tilde{\sigma}_{z_q}$, 30-minute characteristic *e*-folding saltation layer height uncertainty (m)
$z_{q,i}$, characteristic *e*-folding saltation layer height for stress bin $i$ (m)
$\sigma_{z_{q,i}}$, uncertainty in characteristic *e*-folding saltation layer height for stress bin $i$ (m)
$\langle z_q \rangle$, mean saltation layer height for site
$\sigma_{\langle z_q \rangle}$, uncertainty in mean saltation layer height
$\langle z_q \rangle/d_{50}$, mean dimensionless saltation layer height for site
$\sigma_{\langle z_q \rangle/d_{50}}$, uncertainty in mean dimensionless saltation layer height for site
$z_{q,fit,i}$, predicted value for linear best fit of $z_{q,i}$ versus $u_{*,i}$

*Binning variables*
$a$, generic variable name
$\tilde{a}_j$, unbinned 30-minute values for generic variable $a$
$\tilde{\sigma}_{a_j}$, uncertainties for unbinned 30-minute values for generic variable $a$
$a_i$, binned value of generic variable $a$ for stress bin $i$

$\sigma_{a_i}$, uncertainty for binned value of generic variable $a$ for stress bin $i$
$\sigma_{\mu,a_i}$, uncertainty for binned value of generic variable $a$ associated with measurement errors
$\sigma_{SE,a_i}$, uncertainty for binned value of generic variable $a$ associated with bin standard error
$\sigma'_{SE,a_i}$, modified bin standard error uncertainty for binned value of generic variable $a$
$SD_{med,a}$, median standard deviation for binned variable $a$ for bins containing at least 3 values
$N$, total number of stress bins
$N_i$, number of values in stress bin $i$

*Fitting variables*
$y_i$, values of variable for fitting and uncertainty propagation
$\sigma_{y_i}$, uncertainties in values of variable for fitting and uncertainty propagation
$a$, slope parameter for linear fitting or an independent variable for uncertainty propagation
$\sigma_a$, uncertainty in intercept for linear fitting or variable for uncertainty propagation
$b$, intercept parameter for linear fitting or an independent variable for uncertainty propagation
$\sigma_b$, uncertainty in slope for linear fitting or variable for uncertainty propagation
$C$, generic fitting parameter for linear and nonlinear stress-flux relationships (equations S11 and S13)
$\sigma_C$, uncertainty in generic fitting parameter for stress-flux relationships (equations S11 and S13)
$C_Q$, fitting parameter for linear stress-flux relationship (equation 3 in main text)
$\sigma_{C_Q}$, uncertainty in fitting parameter for linear stress-flux relationship (equation 3 in main text)
$\hat{C}_{Q,i}$, estimated dimensionless saltation flux scaling parameter for stress bin $i$
$\sigma_{\hat{C}_{Q,i}}$, uncertainty in estimated dimensionless saltation flux scaling parameter for stress bin $i$
$\langle \hat{C}_Q \rangle$, mean estimated dimensionless saltation flux scaling parameter
$\sigma_{\langle \hat{C}_Q \rangle}$, uncertainty in mean estimated dimensionless saltation flux scaling parameter
$C_t$, alternative fitting parameter for linear stress-flux relationship (equation 12 in main text)
$\sigma_{C_t}$, uncertainty in alternative fitting parameter for linear stress-flux relationship (equation 12 in main text)
$\hat{C}_{t,i}$, estimated alternative dimensionless saltation flux scaling parameter for stress bin $i$
$\sigma_{\hat{C}_{t,i}}$, uncertainty in alternative dimensionless saltation flux scaling parameter for stress bin $i$
$C_z$, constant describing dependence of hop height on saltation layer height (equation 11 in main text)
$\sigma_{u_{*it}}$, uncertainty in fit of impact threshold shear velocity (m s$^{-1}$)
$\sigma_{\tau_{it}}$, uncertainty in fit of impact threshold shear stress (Pa)
$\chi^2$, mean-square difference between observations and predictions
$\chi^2_\nu$, normalized mean-square difference between observations and predictions
$\nu$, degrees of freedom for fit

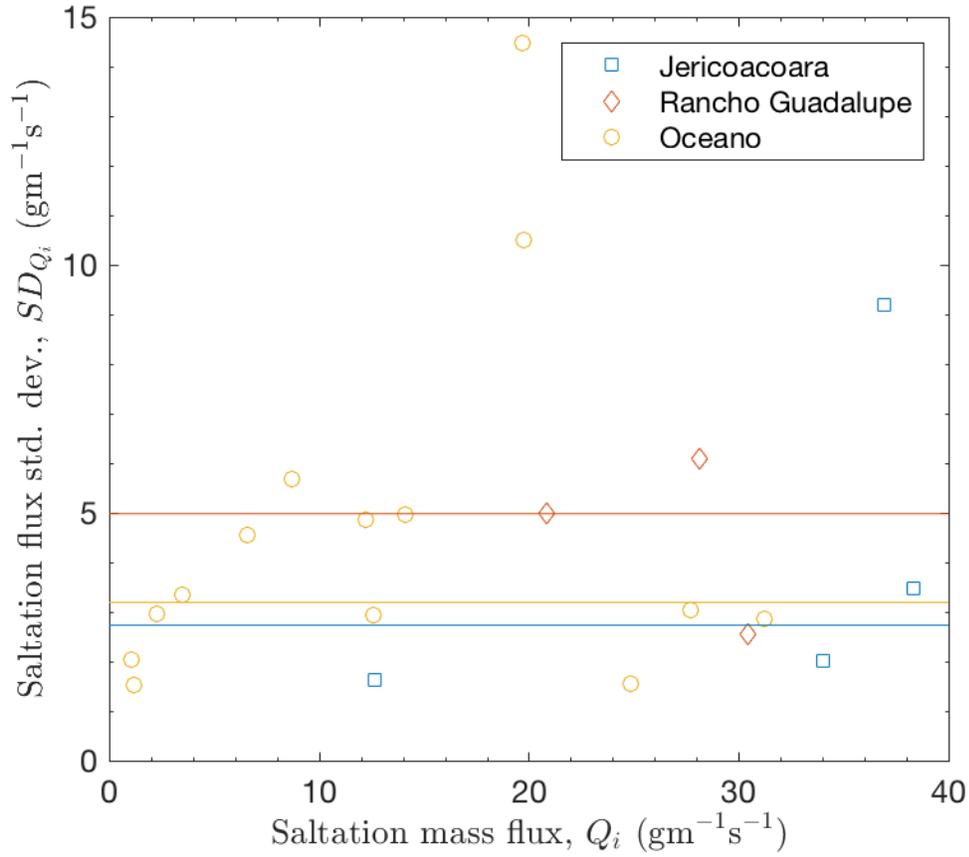

**Fig. S1. Standard deviation of binned values for saltation flux $SD_{Q_i}$ versus binned saltation flux $Q_i$.** Only those bins containing at least 3 values are plotted. Solid lines show resulting medians for these standard deviations $SD_{med,Q}$ for each site. The $SD_{med,Q}$ were then used to calculate standard errors for sparse bins containing fewer than three data points (equation S5).

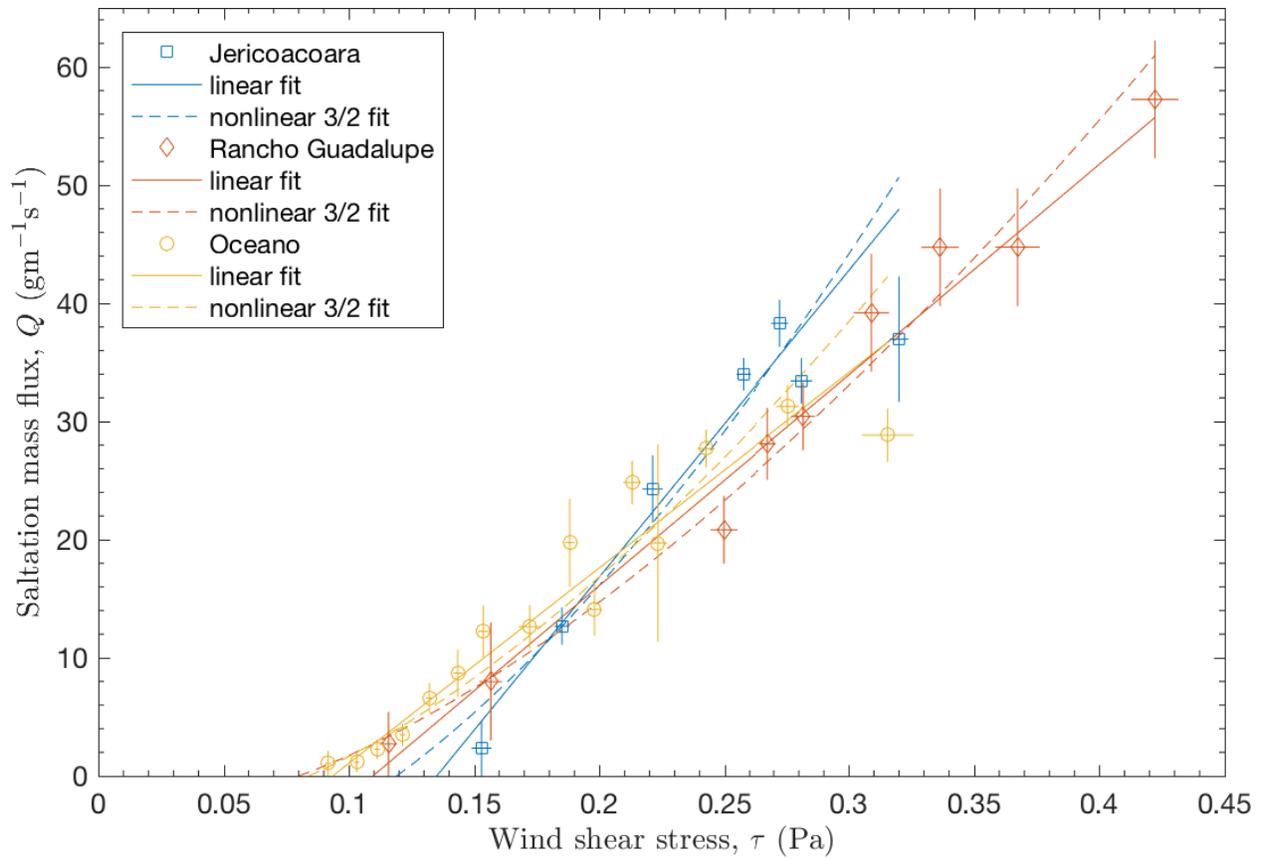

**Fig. S2. Binned saltation flux $Q_i$ versus binned shear stress $\tau_i$.** Error bars indicate uncertainties in binned values for flux ($\sigma_{Q_i}$) and stress ($\sigma_{\tau_i}$). Solid lines show linear fits to equation S11. Dashed lines show nonlinear 3/2 fits to equation S13. Specific parameter values for these fits are listed in Table S1.

**Table S1. Saltation flux law fit values for the three field sites.**

| Site | $d_{50}$ (mm) | Linear: $Q = C(\tau - \tau_{it})$ | | | Nonlinear 3/2: $Q = Cu_*(\tau - \tau_{it})$ | | |
|---|---|---|---|---|---|---|---|
| | | $C$ (s× $10^3$) | $\tau_{it}$ (Pa) | $\chi_\nu^2$ | $C$ (m$^{-1}$s$^2$× $10^3$) | $\tau_{it}$ (Pa) | $\chi_\nu^2$ |
| Jericoacoara | $0.53 \pm 0.04$ | $259 \pm 16$ | $0.135 \pm 0.015$ | 2.59 | $480 \pm 43$ | $0.119 \pm 0.011$ | 3.56 |
| Rancho Guadalupe | $0.53 \pm 0.03$ | $178 \pm 14$ | $0.110 \pm 0.021$ | 0.51 | $303 \pm 35$ | $0.080 \pm 0.025$ | 0.44 |
| Oceano | $0.40 \pm 0.07$ | $165 \pm 6$ | $0.094 \pm 0.006$ | 2.04 | $359 \pm 19$ | $0.084 \pm 0.005$ | 2.62 |

Uncertainties for fits are expressed as $\pm 1$ standard deviation.